\documentclass[runningheads]{llncs}

\usepackage[utf8]{inputenc}

\usepackage{booktabs} % For formal tables

\usepackage{array}
\newcommand{\PreserveBackslash}[1]{\let\temp=\\#1\let\\=\temp}
\newcolumntype{C}[1]{>{\PreserveBackslash\centering}p{#1}}
\newcolumntype{R}[1]{>{\PreserveBackslash\raggedleft}p{#1}}
\newcolumntype{L}[1]{>{\PreserveBackslash\raggedright}p{#1}}

\usepackage{amsmath}
\usepackage{bm}
\usepackage{amsfonts}
\usepackage{multirow}
\usepackage{colortbl}
\usepackage{enumitem}
\usepackage{subcaption}
\usepackage{tabu}
\usepackage{balance}
\usepackage{footnote}
\makesavenoteenv{tabular}
\makesavenoteenv{table}
\usepackage{tikz}
\usepackage{adjustbox}

\usepackage{fontawesome}

\def\eqref#1{equation~\ref{#1}}
\def\1{\bm{1}}

\DeclareMathAlphabet{\mathsfit}{\encodingdefault}{\sfdefault}{m}{sl}
\SetMathAlphabet{\mathsfit}{bold}{\encodingdefault}{\sfdefault}{bx}{n}

\DeclareMathOperator*{\bertcat}{BERT_\text{CAT}}
\DeclareMathOperator*{\bertdot}{BERT_\text{DOT}}

\DeclareMathOperator*{\colbert}{ColBERT}

\usepackage{array, booktabs}
\usepackage{graphicx}

\usepackage{pifont}
\usepackage[noabbrev]{cleveref}

\definecolor{niceGreen1}{RGB}{41, 135, 66}
\definecolor{niceGreen2}{RGB}{147, 184, 103}
\definecolor{niceYellow2}{RGB}{189, 164, 81}
\definecolor{niceRed2}{RGB}{186, 56, 65}

\begin{document}
%\maketitle

\title{Establishing Strong Baselines for~TripClick~Health~Retrieval}

\author{Sebastian Hofst{\"a}tter \and
Sophia Althammer \and
Mete Sertkan \and
Allan Hanbury}
\authorrunning{S. Hofst{\"a}tter et al.}
% First names are abbreviated in the running head.
% If there are more than two authors, 'et al.' is used.
%
\institute{TU Wien, Austria\\
\email{\{first.last\}@tuwien.ac.at}}

\maketitle
\begin{abstract}
\vspace{-0.7cm}
We present strong Transformer-based re-ranking and dense retrieval baselines for the recently released TripClick health ad-hoc retrieval collection. We improve the -- originally too noisy -- training data with a simple negative sampling policy. We achieve large gains over BM25 in the re-ranking task of TripClick, which were not achieved with the original baselines. Furthermore, we study the impact of different domain-specific pre-trained models on TripClick. Finally, we show that dense retrieval outperforms BM25 by considerable margins, even with simple training procedures. 
\vspace{-0.5cm}
\end{abstract}

\section{Introduction}
\vspace{-0.2cm}
The latest neural network advances in Information Retrieval (IR) -- specifically the ad-hoc passage retrieval task -- are driven by available training data, especially the large web-search-based MSMARCO collection \cite{msmarco16}. Here, neural approaches lead to enormous effectiveness gains over traditional techniques \cite{hofstaetter2020_crossarchitecture_kd,khattab2020colbert,macavaney2019,nogueira2019passage}. 
A valid concern is the generalizability and applicability of the developed techniques to other domains and settings \cite{lin2021proposed,lietal-multitask21,thakur2021beir,wang-2021-TSDAE}. 

The newly released TripClick collection \cite{rekabsaz2021tripclick} with large-scale click log data from the \textit{Trip Database}, a health search engine, provides us with the opportunity to re-test previously developed techniques on this new ad-hoc retrieval task: keyword search in the health domain with large training and evaluation sets. TripClick provides three different test sets (Head, Torso, Tail), grouped by their query frequency, so we can analyze model performance for different slices of the overall query distribution.

This study conducts a range of controlled ad-hoc retrieval experiments using pre-trained Transformer \cite{vaswani2017attention} models with various state-of-the-art retrieval architectures on the TripClick collection. We aim to reproduce effectiveness gains achieved on MSMARCO in the click-based health ad-hoc retrieval setting. Typically, neural ranking models are trained with a triple of one query, a relevant and a non-relevant passage. As part of our evaluation study, we discovered a flaw in the provided neural training data of TripClick: The original negative sampling strategy included non-clicked results, which led to inadequate training. Therefore, we re-created the training data with an improved negative sampling strategy, based solely on BM25 negatives, with better results than published baselines.

As the TripClick collection was released only recently, we are the first to study a wide-ranging number of BERT-style ranking architectures and answer the fundamental question:

\newcommand{\RQone}{\begin{itemize}
    \item[\textbf{RQ1}] How do established ranking models perform on re-ranking TripClick?
\end{itemize}}
\newcommand{\RQoneRunning}{\textbf{RQ1} \textit{How do established ranking models perform on re-ranking TripClick? }}
\RQone

In the re-ranking setting, where the neural models score a set of $200$ candidates produced by BM25, we observe large effectiveness gains for $\bertcat$, $\colbert$, and TK for every one of the three frequency-based query splits. $\bertcat$ improves over BM25 on Head by $100\%$, on Torso by $66\%$ and Tail still by $50\%$.

We compare the general BERT-Base \& DistilBERT with the domain-specific SciBERT \& PubMedBERT models to answer:
\newcommand{\RQtwo}{\begin{itemize}
    \item[\textbf{RQ2}] Which BERT-style pre-trained checkpoint performs best on TripClick?
\end{itemize}}
\newcommand{\RQtwoRunning}{\textbf{RQ2} \textit{Which BERT-style pre-trained checkpoint performs best on TripClick? }}

\RQtwo

Although the general domain models show good effectiveness results, they are outperformed by the domain-specific pre-training approaches. Here, PubMedBERT slightly outperforms SciBERT on re-ranking with $\bertcat$ \& $\colbert$. An ensemble of all domain-specific models with $\bertcat$ again outperforms all previous approaches and sets new state-of-the-art results for TripClick.

Finally, we study the concept of retrieving passages directly from a nearest neighbor vector index, also referred to as dense retrieval, and answer:
\newcommand{\RQthree}{\begin{itemize}
    \item[\textbf{RQ3}] How well does dense retrieval work on TripClick?
\end{itemize}}
\newcommand{\RQthreeRunning}{\textbf{RQ3} \textit{How well does dense retrieval work on TripClick? }}

\RQthree

Dense retrieval outperforms BM25 considerably for initial candidate retrieval, both in top-10 precision results and for all recall cutoffs, except top-1000. In contrast to re-ranking, SciBERT outperforms PuBMedBERT on dense retrieval results.

We publish our source code as well as the improved training triples at: \\ \url{\textit{https://github.com/sebastian-hofstaetter/tripclick}}

\vspace{-0.2cm}
\section{Background}
\vspace{-0.2cm}
We describe the collection, the BERT-style pre-training instances, ranking architectures, and training procedures we use below.

\vspace{-0.2cm}
\subsection{TripClick Collection}

TripClick contains $1.5$ million passages (with an average length of $259$ words), $680$ thousand click-based training queries (with an average of $4.4$ words), and $3,525$ test queries. The TripClick collection includes three test sets with $1,175$ queries each grouped by their frequency and called Head, Torso, and Tail queries. For the Head queries a DCTR \cite{chuklin2015click} click model was employed to created relevance signals, the other two sets use raw clicks. 

In comparison to the widely analyzed MSMARCO collection \cite{hofstaetter2021mitigating}, TripClick is yet to be fully understood. This includes the quality of the click labels and the effect of various filtering mechanisms of the professional search production UI, that are not part of the released data.\footnote{The TripDatabase allows users to use different ranking schemes, such as popularity, source quality and pure relevance, as well as filtering results by facets. Unfortunately, this information is not available in the public dataset.}
\subsection{Re-ranking \& Retrieval Models}

We study multiple architectures with different aspects on the efficiency vs. effectiveness tradeoff scale. Here, we give a brief overview, for more detailed comparisons see Hofst{\"a}tter et al. \cite{hofstaetter2020_crossarchitecture_kd}.

\vspace{0.1cm}
\noindent{\textbf{BERT$_\textbf{CAT}$ -- Concatenated Scoring}} The base re-ranking model BERT$_\text{CAT}$ \cite{nogueira2019passage,macavaney2019,yilmaz2019cross} concatenates query and passage sequences with special tokens and computes a score by reducing the pooled CLS representation with a single linear layer.
It represents one of the current state-of-the art models in terms of effectiveness, however it exhibits many drawbacks in terms of efficiency \cite{Hofstaetter2019_osirrc,xiong2020approximate}.

\vspace{0.1cm}
\noindent{\textbf{ColBERT}} The $\colbert$ model \cite{khattab2020colbert} delays the interactions between every query and document representation after BERT.
The interactions in the $\colbert$ model are aggregated with a max-pooling per query term and sum of query-term scores.
The aggregation only requires simple dot product computations, however the storage cost of pre-computing passage representations is very high as it depends on the total number of terms in the collection.

\vspace{0.1cm}

\noindent{\textbf{TK (Transformer-Kernel)}} The Transformer-Kernel model \cite{Hofstaetter2020_ecai} is not based on BERT pre-training, but rather uses shallow and independently computed Transformers followed by a set of RBF kernels to count match signals in a term-by-term match matrix, for very efficient re-ranking. 

\vspace{0.1cm}
\noindent{\textbf{BERT{$_\textbf{DOT} $} -- Dense Retrieval}} The BERT$_\text{DOT}$ model matches a single CLS vector of the query with a single CLS vector of a passage \cite{xiong2020approximate,luan2020sparse,lu2020twinbert}, independently computed.
This decomposition of interactions to a single dot-product allows us to pre-compute every contextualized passage representation and employ a nearest neighbor index for dense retrieval, without a traditional first stage.

\vspace{-0.2cm}
\subsection{Pre-trained BERT Instances}

The 12-layer BERT-Base model \cite{devlin2018bert} (and the 6-layer distilled version DistilBERT \cite{sanh2019distilbert}) and its vocabulary are based on the Books Corpus and English Wikipedia articles. The SciBERT model \cite{beltagy-etal-2019-scibert} uses the identical architecture to the BERT-Base model, but the vocabulary and the weights are pre-trained on Semantic Scholar articles (with $82\%$ articles from the broad biomedical domain). Similarly the PubMedBERT model \cite{pubmedbert2020} and its vocabulary are trained on PubMed articles using the same architecture as the BERT model.

\vspace{-0.2cm}
\subsection{Related Studies}

At the time of writing, this is the first paper evaluating on the novel TripClick collection. However many other tasks have been set up before in the biomedical retrieval domain, such as BioASQ \cite{bioasq}, TREC Precision Medicine tracks \cite{trecprecmedicine,trechealthmisinformation} or the timely created TREC-COVID \cite{treccovid,moller-etal-2020-covid,covidjimmylin} (which is based on CORD-19 \cite{wang2020cord19}, a collection of scientific articles concerned with the coronavirus pandemic). 

For TREC-COVID, MacAvaney et al. \cite{macavaney2020sledge} train a neural re-ranking model on a subset of the MS MARCO dataset containing only medical terms (Med-MARCO) and demonstrate its domain-focused effectiveness on a transfer to TREC-COVID.
Xiong et al. \cite{xiong2020cmttreccovid} and Lima et al. \cite{lima2020denmarkstreccovid} explore medical domain specific BERT representations for the retrieval from the TREC-COVID corpus and show that using SciBERT for dense retrieval outperforms the BM25 baseline by a large margin. Wang et al. \cite{wang2020participation} explore continuous active learning for the retrieval task from the COVID-19 corpus, this method is also studied for retrieval in the precision medicine track \cite{trecprecmedicine,Cormack_Grossman_2018a}. Reddy et al. \cite{reddy2020endtoend} demonstrate synthetic training for question answering of COVID-19 related questions.

Many of these related works are concerned with overcoming the lack of large training data on previous medical collections. Now with TripClick we have a large-scale medical retrieval dataset. In this paper we jumpstart work on this collection, by showcasing the effectiveness of neural ranking approaches on TripClick.

\vspace{-0.2cm}
\section{Experiment Design}
\label{sec:results}
\vspace{-0.2cm}

In our experiment setup, we largely follow Hofst{\"a}tter et al.  \cite{hofstaetter2020_crossarchitecture_kd}, except where noted otherwise. Mainly we rely on PyTorch \cite{pytorch2017} and HuggingFace Transformer \cite{wolf2019huggingface} libraries as foundation for our neural training and evaluation methods. For TK, we follow Rekabsaz et al. \cite{rekabsaz2021tripclick} and utilize a PubMed-trained $400$ dimensional word embedding as starting point \cite{mcdonald2018deep}.
For validation and testing we utilize the data splits outlined in TripClick by Rekabsaz et al. \cite{rekabsaz2021tripclick}.

\vspace{-0.2cm}
\subsection{Training Data Generation}

The TripClick dataset conveniently comes with a set of pre-generated training triples for neural training. Nevertheless, we found this training set to produce less than optimal results and the trained BERT models show no robustness against increased re-ranking depth. This phenomena of having to tune the best re-ranking depth for effectiveness, rather than efficiency, has been studied as part of early non-BERT re-rankers \cite{Hofstaetter2019_sigir}. With the advent of Transformer-based re-rankers, this technique became obsolete \cite{Hofstaetter2020_ecai}.

In the TripClick dataset, the clicked results are considered as positives samples for training.
However, we discovered a flaw in the published negative sampling procedure, that non-clicked results -- ranked above the clicked ones -- are included as negative sampled passages. We hypothesize this leads to many false negatives in the training set, confusing the models during training. We confirm this thesis by observing our training telemetry data, showing low pairwise training accuracy as well as a lack of clear distinction in the scoring margins of the $\bertcat$ models. For all results presented in this study we generate new training data with the following simple procedure: 
\begin{enumerate}
    \item We generate $500$ BM25 candidates for every training query
    \item For every pair of query - relevant (clicked) passage in the training set we randomly sample, without replacement, up to $20$ negative candidates from the candidates created in 1.
    \begin{itemize}
        \item We remove candidates present in the relevant pool, regardless of relevance grade.
        \item We discard positional information (we expect position bias to be in the training data -- a potential for future work).
    \end{itemize}
    \item After shuffling the training triples we save $10$ million triples for training
\end{enumerate}

Our new training set gave us a $45-50\%$ improvement on MRR@10 (from $.41$ to $.6$) and nDCG@10 (from $.21$ to $.30$) for the HEAD validation queries using the same PubMedBERT$_\text{CAT}$ model and setup. The models are now also robust against increasing the re-ranking depth.

\begin{table*}[t!]
    \centering
    \caption{Effectiveness results for the three frequency-binned TripClick query sets. \textit{The nDCG \& MRR cutoff is at rank 10.}}
    \label{tab:rerank_results}
    %\vspace{-0.3cm}
    \setlength\tabcolsep{1.5pt}
    %\small %L{0.3cm}
    \begin{tabular}{cll!{\color{lightgray}\vrule}rr!{\color{lightgray}\vrule}rr!{\color{lightgray}\vrule}rr}
       \toprule
       \multirow{2}{*}{} & \multirow{2}{*}{\textbf{Model}} & \textbf{BERT} &
       \multicolumn{2}{c!{\color{lightgray}\vrule}}{\textbf{Head} \tiny{(DCTR)}}&
       \multicolumn{2}{c!{\color{lightgray}\vrule}}{\textbf{Torso} \tiny{(RAW)}}&
       \multicolumn{2}{c}{\textbf{Tail} \tiny{(RAW)}}\\
       &&\textbf{Instance}& \small{nDCG} & \small{MRR}  & \small{nDCG} & \small{MRR} &  \small{nDCG} & \small{MRR}  \\
        \midrule
        \multicolumn{6}{l}{\textbf{Original Baselines}} \\
         \textcolor{gray}{1} & BM25      & -- &   .140 & .276 &  .206 & .283 &  .267 & .258  \\

         \textcolor{gray}{2} & ConvKNRM     & -- &  .198 & .420 & .243 & .347  & .271 & .265 \\
         \textcolor{gray}{3} & TK   & -- &   .208 & .434 &  .272 & .381  & .295 & .280 \\

        \midrule
         \multicolumn{6}{l}{\textbf{Our Re-Ranking (BM25 Top-200)}} \\
         \arrayrulecolor{lightgray}

         \textcolor{gray}{4} & \multirow{1}{*}{TK}                 & --           &  .232 & .472  & .300 & .390 & .345 & .319  \\
         \midrule

         \textcolor{gray}{5} &\multirow{2}{*}{ColBERT}             & \multirow{1}{*}{SciBERT}  &  .270 & .556 & .326 & .426 & .374  & .347  \\
         \textcolor{gray}{6} &                                      & \multirow{1}{*}{PubMedBERT-Abstract}   &  .278 & .557 & .340 & .431 & .387 & .361  \\
         \midrule

         \textcolor{gray}{7} &\multirow{6}{*}{BERT$_\text{CAT}$}   & DistilBERT   &  .272 & .556 & .333 & .427 & .381 & .355   \\
         \textcolor{gray}{8} &                                     & BERT-Base     &  .287 & .579 & .349 & .453 & .396 & .366   \\
         \cmidrule{3-9}
         \textcolor{gray}{9} &                                     & SciBERT       &  .294 & .595 & .360 & .459 & .408 & .377  \\
         \textcolor{gray}{10} &                                     & PubMedBERT-Full  &  .298 & .582 & .365 & .462 & .412 & .381  \\
         \textcolor{gray}{11} &                                     & PubMedBERT-Abstract    &  .296 & .587 & .359 & .456 & .409 & .380  \\
         \cmidrule{3-9}
         \textcolor{gray}{12} &                       & 

         \textit{Ensemble (Lines: 9,10,11)}   &  \textbf{.303} & \textbf{.601} & \textbf{.370} & \textbf{.472} & \textbf{.420} & \textbf{.392}  \\

        \arrayrulecolor{black}
        \bottomrule
    \end{tabular}
    \vspace{-0.3cm}
\end{table*}

\vspace{-0.2cm}
\section{TripClick Effectiveness Results}
\label{sec:results}
\vspace{-0.2cm}

In this section, we present the results for our research questions, first for re-ranking and then for dense retrieval.

\vspace{-0.2cm}
\subsection{Re-Ranking}
\vspace{-0.2cm}

We present the original baselines, as well as our re-ranking results for all three frequency-based TripClick query sets in Table \ref{tab:rerank_results}. All neural models re-rank the top-200 results of BM25. While the original baselines do improve the frequent Head queries by up to $6$ points nDCG@10 (TK-L3 vs. BM25-L1); they hardly improve the Tail queries with only $1-3$ points difference in nDCG@10 (CK-L2 \& TK-L3 vs. BM25-L1). This is a pressing issue, as those queries make up 83\% of all Trip searches \cite{rekabsaz2021tripclick}. 

Turning to our results in Table \ref{tab:rerank_results}, to answer \RQoneRunning We can see that our training approach for TK (Line 4) strongly outperforms the original TK (L3), especially on the Tail queries. This is followed by ColBERT (L5 \& 6) and $\bertcat$ (L7 to L12) which both improve strongly over the previous model. This trend directly follows previous observations of effectiveness improvements per model architecture on MSMARCO \cite{hofstaetter2020_crossarchitecture_kd,khattab2020colbert}.

To understand if there is a clear benefit of the BERT model choice we study: \RQtwoRunning We find that although the general domain models show good effectiveness results (L7 \& 8), they are outperformed by the domain-specific pre-training approaches (L9 to L11). Here, PubMedBERT (L5 + L10 \& 11) slightly outperforms SciBERT (L9 + L10 \& 11) on re-ranking with $\bertcat$ \& $\colbert$. An ensemble of all domain-specific models with $\bertcat$ (L12) again outperforms all previous approaches, and sets new state-of-the-art results for TripClick.

\vspace{-0.2cm}
\subsection{Dense Retrieval}

To answer \RQthreeRunning we present our results in Table \ref{tab:dense_results}. Dense retrieval with $\bertdot$ (L13 to L15) outperforms BM25 (L1) considerably for initial candidate retrieval, both in terms of top-10 precision results, as well as for all recall cutoffs, except top-1000. We also provided the judgement coverage for the top-10 results, and surprisingly, the coverage for dense retrieval increases compared to BM25. Future annotation campaigns should explore the robustness of these click-based evaluation results.

\begin{table}[t!]
    \centering
    \caption{BERT$_\textbf{DOT}$ dense retrieval effectiveness results for the HEAD TripClick query set. \textit{J@10 indicates the ratio of judged results at cutoff 10.}}
    \label{tab:dense_results}
    %\vspace{-0.3cm}
    \setlength\tabcolsep{2pt}
    %\small %L{0.3cm}
    \begin{tabular}{cll!{\color{lightgray}\vrule}c!{\color{lightgray}\vrule}rr!{\color{lightgray}\vrule}rrr}
       \toprule
       \multirow{2}{*}{} & \multirow{2}{*}{\textbf{Model}}
       & %\textbf{BERT}& 
       \multirow{1}{*}{\textbf{BERT}}   &
       \multicolumn{5}{c}{\textbf{Head} (DCTR)}\\
       &&\textbf{Instance}& \small{J@10} & \small{nDCG@10} & \small{MRR@10}  & \small{R@100} & \small{R@200} & \small{R@1K}  \\
        \midrule
        \multicolumn{6}{l}{\textbf{Original Baselines}} \\
         \textcolor{gray}{1} & BM25      & -- & 31\% & .140 & .276 & .499 & .621 & \textbf{.834} \\

        \arrayrulecolor{black}
        \midrule
        \arrayrulecolor{lightgray}
        \multicolumn{9}{l}{\textbf{Retrieval (Full Collection Nearest Neighbor)}} \\
        % multirow{8}{*}{BERT$_\text{DOT}$}
         \textcolor{gray}{13} & \multirow{3}{*}{BERT$_\text{DOT}$} & \multirow{1}{*}{DistilBERT}   & 39\% & .236 & .512 & .550 & .648 & .813  \\
         \textcolor{gray}{14}                                      & &\multirow{1}{*}{SciBERT} & 41\% & \textbf{.243} & \textbf{.530} & .562 & .640 & .793   \\
         \textcolor{gray}{15}                                      & &\multirow{1}{*}{PubMedBERT}   & 40\%&  .235 & .509 & \textbf{.582} & \textbf{.673} & .828  \\
         
        \arrayrulecolor{black}
        
        \arrayrulecolor{black}
        \bottomrule
    \end{tabular}
    \vspace{-0.3cm}
\end{table}

\vspace{-0.2cm}
\section{Conclusion}
\vspace{-0.2cm}

Test collection diversity is a fundamental requirement of IR research. Ideally, we as a community develop methods that work on the largest possible set of problem settings. However, neural models require large training sets, which restricted most of the foundational research to the public MSMARCO and other web search collections. Now, with TripClick we have a another large-scale collection available. In this paper we show that in contrast to the original baselines, neural models perform very well on TripClick -- both in the re-ranking task and the full collection retrieval with nearest neighbor search. We make our techniques openly available to the community to foster diverse neural information retrieval research.

%\newpage
%\balance
\balance
\bibliographystyle{abbrv}

\bibliography{references}

\begin{thebibliography}{10}

\bibitem{msmarco16}
P.~Bajaj, D.~Campos, N.~Craswell, L.~Deng, J.~Gao, X.~Liu, R.~Majumder,
  A.~Mcnamara, B.~Mitra, and T.~Nguyen.
\newblock {MS MARCO : A Human Generated MAchine Reading COmprehension Dataset}.
\newblock In {\em Proc. of NIPS}, 2016.

\bibitem{beltagy-etal-2019-scibert}
I.~Beltagy, K.~Lo, and A.~Cohan.
\newblock {S}ci{BERT}: A pretrained language model for scientific text.
\newblock In {\em Proc. of EMNLP-IJCNLP}, 2019.

\bibitem{chuklin2015click}
A.~Chuklin, I.~Markov, and M.~de~Rijke.
\newblock {\em Click Models for Web Search}.
\newblock Morgan \& Claypool, 2015.

\bibitem{Cormack_Grossman_2018a}
G.~Cormack and M.~Grossman.
\newblock Technology-assisted review in empirical medicine: Waterloo
  participation in clef ehealth 2018.
\newblock In {\em CLEF (Working Notes)}, 2018.

\bibitem{devlin2018bert}
J.~Devlin, M.~Chang, K.~Lee, and K.~Toutanova.
\newblock Bert: Pre-training of deep bidirectional transformers for language
  understanding.
\newblock In {\em Proc. of NAACL}, 2019.

\bibitem{trechealthmisinformation}
M.~Fern{\'a}ndez-Pichel, D.~Losada, J.~C. Pichel, and D.~Elsweiler.
\newblock Citius at the trec 2020 health misinformation track.
\newblock 2020.

\bibitem{pubmedbert2020}
Y.~Gu, R.~Tinn, H.~Cheng, M.~Lucas, N.~Usuyama, X.~Liu, T.~Naumann, J.~Gao, and
  H.~Poon.
\newblock Domain-specific language model pretraining for biomedical natural
  language processing.
\newblock 2020.

\bibitem{hofstaetter2020_crossarchitecture_kd}
S.~Hofst{\"a}tter, S.~Althammer, M.~Schr{\"o}der, M.~Sertkan, and A.~Hanbury.
\newblock Improving efficient neural ranking models with cross-architecture
  knowledge distillation.
\newblock {\em arXiv preprint2010.02666}, 2020.

\bibitem{Hofstaetter2019_osirrc}
S.~Hofst{\"a}tter and A.~Hanbury.
\newblock {Let's measure run time! Extending the IR replicability
  infrastructure to include performance aspects}.
\newblock In {\em Proc. of OSIRRC}, 2019.

\bibitem{hofstaetter2021mitigating}
S.~Hofst{\"a}tter, A.~Lipani, S.~Althammer, M.~Zlabinger, and A.~Hanbury.
\newblock Mitigating the position bias of transformer models in passage
  re-ranking.
\newblock In {\em Proc. of ECIR}, 2021.

\bibitem{Hofstaetter2019_sigir}
S.~Hofst{\"a}tter, N.~Rekabsaz, C.~Eickhoff, and A.~Hanbury.
\newblock {On the Effect of Low-Frequency Terms on Neural-IR Models}.
\newblock In {\em Proc. of SIGIR}, 2019.

\bibitem{Hofstaetter2020_ecai}
S.~Hofst{\"a}tter, M.~Zlabinger, and A.~Hanbury.
\newblock {Interpretable \& Time-Budget-Constrained Contextualization for
  Re-Ranking}.
\newblock In {\em Proc. of ECAI}, 2020.

\bibitem{khattab2020colbert}
O.~Khattab and M.~Zaharia.
\newblock Colbert: Efficient and effective passage search via contextualized
  late interaction over bert.
\newblock In {\em Proc. of SIGIR}, 2020.

\bibitem{lietal-multitask21}
M.~Li, M.~Li, K.~Xiong, and J.~Lin.
\newblock Multi-task dense retrieval via model uncertainty fusion for
  open-domain question answering.
\newblock In {\em Findings of EMNLP}, 2021.

\bibitem{lima2020denmarkstreccovid}
L.~C. Lima, C.~Hansen, C.~Hansen, D.~Wang, M.~Maistro, B.~Larsen, J.~G.
  Simonsen, and C.~Lioma.
\newblock Denmark's participation in the search engine trec covid-19 challenge:
  Lessons learned about searching for precise biomedical scientific information
  on covid-19.
\newblock {\em arXiv preprint2011.12684}, 2020.

\bibitem{lin2021proposed}
J.~Lin.
\newblock A proposed conceptual framework for a representational approach to
  information retrieval.
\newblock {\em arXiv preprint2110.01529}, 2021.

\bibitem{lu2020twinbert}
W.~Lu, J.~Jiao, and R.~Zhang.
\newblock Twinbert: Distilling knowledge to twin-structured bert models for
  efficient retrieval.
\newblock {\em arXiv preprint arXiv:2002.06275}, 2020.

\bibitem{luan2020sparse}
Y.~Luan, J.~Eisenstein, K.~Toutanova, and M.~Collins.
\newblock Sparse, dense, and attentional representations for text retrieval.
\newblock {\em arXiv preprint arXiv:2005.00181}, 2020.

\bibitem{macavaney2020sledge}
S.~MacAvaney, A.~Cohan, and N.~Goharian.
\newblock Sledge: A simple yet effective baseline for covid-19 scientific
  knowledge search.
\newblock {\em arXiv preprint2005.02365}, 2020.

\bibitem{macavaney2019}
S.~MacAvaney, A.~Yates, A.~Cohan, and N.~Goharian.
\newblock Cedr: Contextualized embeddings for document ranking.
\newblock In {\em Proc. of SIGIR}, 2019.

\bibitem{mcdonald2018deep}
R.~McDonald, G.-I. Brokos, and I.~Androutsopoulos.
\newblock Deep relevance ranking using enhanced document-query interactions.
\newblock {\em arXiv preprint1809.01682}, 2018.

\bibitem{moller-etal-2020-covid}
T.~M{\"o}ller, A.~Reina, R.~Jayakumar, and M.~Pietsch.
\newblock {COVID-QA}: A question answering dataset for {COVID}-19.
\newblock In {\em Proceedings of the 1st Workshop on {NLP} for {COVID-19} at
  {ACL} 2020}, Online, July 2020. Association for Computational Linguistics.

\bibitem{bioasq}
A.~Nentidis, A.~Krithara, K.~Bougiatiotis, M.~Krallinger, C.~Rodriguez-Penagos,
  M.~Villegas, and G.~Paliouras.
\newblock {\em Overview of BioASQ 2020: The Eighth BioASQ Challenge on
  Large-Scale Biomedical Semantic Indexing and Question Answering}, pages
  194--214.
\newblock 09 2020.

\bibitem{nogueira2019passage}
R.~Nogueira and K.~Cho.
\newblock Passage re-ranking with bert.
\newblock {\em arXiv preprint arXiv:1901.04085}, 2019.

\bibitem{pytorch2017}
A.~Paszke, S.~Gross, S.~Chintala, G.~Chanan, E.~Yang, Z.~DeVito, Z.~Lin,
  A.~Desmaison, L.~Antiga, and A.~Lerer.
\newblock Automatic differentiation in pytorch.
\newblock In {\em Proc. of NIPS-W}, 2017.

\bibitem{reddy2020endtoend}
R.~G. Reddy, B.~Iyer, M.~A. Sultan, R.~Zhang, A.~Sil, V.~Castelli, R.~Florian,
  and S.~Roukos.
\newblock End-to-end qa on covid-19: Domain adaptation with synthetic training.
\newblock {\em arXiv preprint2012.01414}, 2020.

\bibitem{rekabsaz2021tripclick}
N.~Rekabsaz, O.~Lesota, M.~Schedl, J.~Brassey, and C.~Eickhoff.
\newblock Tripclick: The log files of a large health web search engine.
\newblock {\em arXiv preprint2103.07901}, 2021.

\bibitem{trecprecmedicine}
K.~Roberts, D.~Demner-Fushman, E.~Voorhees, W.~Hersh, S.~Bedrick, A.~J. Lazar,
  and S.~Pant.
\newblock Overview of the trec 2019 precision medicine track.
\newblock {\em The ... text REtrieval conference : TREC. Text REtrieval
  Conference}, 26, 2019.

\bibitem{sanh2019distilbert}
V.~Sanh, L.~Debut, J.~Chaumond, and T.~Wolf.
\newblock Distilbert, a distilled version of bert: smaller, faster, cheaper and
  lighter.
\newblock {\em arXiv preprint arXiv:1910.01108}, 2019.

\bibitem{covidjimmylin}
R.~Tang, R.~Nogueira, E.~Zhang, N.~Gupta, P.~Cam, K.~Cho, and J.~Lin.
\newblock Rapidly bootstrapping a question answering dataset for covid-19.
\newblock {\em CoRR}, abs/2004.11339, 2020.

\bibitem{thakur2021beir}
N.~Thakur, N.~Reimers, A.~Rücklé, A.~Srivastava, and I.~Gurevych.
\newblock Beir: A heterogenous benchmark for zero-shot evaluation of
  information retrieval models.
\newblock {\em arXiv preprint arXiv:2104.08663}, 4 2021.

\bibitem{vaswani2017attention}
A.~Vaswani, N.~Shazeer, N.~Parmar, J.~Uszkoreit, et~al.
\newblock Attention is all you need.
\newblock In {\em Proc. of NIPS}, 2017.

\bibitem{treccovid}
E.~Voorhees, T.~Alam, S.~Bedrick, D.~Demner-Fushman, W.~Hersh, K.~Lo,
  K.~Roberts, I.~Soboroff, and L.~L. Wang.
\newblock Trec-covid: Constructing a pandemic information retrieval test
  collection.
\newblock {\em ArXiv}, abs/2005.04474, 2020.

\bibitem{wang-2021-TSDAE}
K.~Wang, N.~Reimers, and I.~Gurevych.
\newblock Tsdae: Using transformer-based sequential denoising auto-encoderfor
  unsupervised sentence embedding learning.
\newblock {\em arXiv preprint arXiv:2104.06979}, 4 2021.

\bibitem{wang2020cord19}
L.~L. Wang, K.~Lo, Y.~Chandrasekhar, R.~Reas, J.~Yang, D.~Burdick, D.~Eide,
  K.~Funk, Y.~Katsis, R.~Kinney, Y.~Li, Z.~Liu, W.~Merrill, P.~Mooney,
  D.~Murdick, D.~Rishi, J.~Sheehan, Z.~Shen, B.~Stilson, A.~Wade, K.~Wang,
  N.~X.~R. Wang, C.~Wilhelm, B.~Xie, D.~Raymond, D.~S. Weld, O.~Etzioni, and
  S.~Kohlmeier.
\newblock Cord-19: The covid-19 open research dataset.
\newblock {\em arXiv preprint2004.10706}, 2020.

\bibitem{wang2020participation}
X.~J. Wang, M.~R. Grossman, and S.~G. Hyun.
\newblock Participation in trec 2020 covid track using continuous active
  learning.
\newblock {\em arXiv preprint2011.01453}, 2020.

\bibitem{wolf2019huggingface}
T.~Wolf, L.~Debut, V.~Sanh, J.~Chaumond, C.~Delangue, A.~Moi, P.~Cistac,
  T.~Rault, R.~Louf, M.~Funtowicz, et~al.
\newblock Huggingface's transformers: State-of-the-art natural language
  processing.
\newblock {\em ArXiv}, pages arXiv--1910, 2019.

\bibitem{xiong2020cmttreccovid}
C.~Xiong, Z.~Liu, S.~Sun, Z.~Dai, K.~Zhang, S.~Yu, Z.~Liu, H.~Poon, J.~Gao, and
  P.~Bennett.
\newblock Cmt in trec-covid round 2: Mitigating the generalization gaps from
  web to special domain search.
\newblock {\em arXiv preprint2011.01580}, 2020.

\bibitem{xiong2020approximate}
L.~Xiong, C.~Xiong, Y.~Li, K.-F. Tang, J.~Liu, P.~Bennett, J.~Ahmed, and
  A.~Overwijk.
\newblock Approximate nearest neighbor negative contrastive learning for dense
  text retrieval.
\newblock {\em arXiv preprint arXiv:2007.00808}, 2020.

\bibitem{yilmaz2019cross}
Z.~A. Yilmaz, W.~Yang, H.~Zhang, and J.~Lin.
\newblock Cross-domain modeling of sentence-level evidence for document
  retrieval.
\newblock In {\em Proc. of EMNLP-IJCNLP}, 2019.

\end{thebibliography}

\end{document}